\documentclass[a4paper,11pt,oneside]{article}

\usepackage[english]{babel}
\usepackage[T1]{fontenc}
\usepackage[utf8]{inputenc}

\usepackage[pdftex,unicode]{hyperref}
\hypersetup{pdftitle=Ranking-Based Second Stage in Data Envelopment Analysis: An Application to Research Efficiency in Higher Education}
\hypersetup{pdfauthor=Vladimír Holý}

\usepackage{xcolor}
\definecolor{mycolor}{rgb}{0.50,0.25,0.50}
\hypersetup{colorlinks=true, linkcolor=mycolor, anchorcolor=mycolor, citecolor=mycolor, filecolor=mycolor, urlcolor=mycolor}

\usepackage[margin=60pt]{geometry}
\usepackage{amsmath}
\usepackage{amssymb}
\usepackage{amsthm}

\usepackage[group-minimum-digits=3]{siunitx}

\usepackage{graphicx}
\usepackage{float}
\usepackage{hhline}
\usepackage{multirow}
\usepackage{booktabs}
\usepackage{dcolumn}
\usepackage{pdflscape}
\usepackage{array}

\usepackage[authoryear]{natbib}

\begin{document}

\begin{center}
{\Large \bfseries Ranking-Based Second Stage in Data Envelopment Analysis: \\ An Application to Research Efficiency in Higher Education}
\end{center}

\begin{center}
{\bfseries Vladimír Holý} \\
Prague University of Economics and Business \\
Winston Churchill Square 1938/4, 130 67 Prague 3, Czechia \\
\href{mailto:vladimir.holy@vse.cz}{vladimir.holy@vse.cz} \\
\end{center}

\noindent
\textbf{Abstract:}
An alternative approach for the panel second stage of data envelopment analysis (DEA) is presented in this paper. Instead of efficiency scores, we propose to model rankings in the second stage using a dynamic ranking model in the score-driven framework. We argue that this approach is suitable to complement traditional panel regression as a robustness check. To demonstrate the proposed approach, we determine research efficiency of higher education systems at country level by examining scientific publications and analyze its relation to good governance. The proposed approach confirms positive relation to the Voice and Accountability indicator, as found by the standard panel linear regression, while suggesting caution regarding the Government Effectiveness indicator.
\\

\noindent
\textbf{Keywords:} Two-Stage DEA, Ranking Model, Score-Driven Model, Research \& Development.
\\

\noindent
\textbf{JEL Classification:} C33, C44, I23, O32.
\\

\section{Introduction}
\label{sec:intro}

In operations research, data envelopment analysis (DEA) is a non-parametric method used to measure the relative efficiency of decision-making units (DMUs) that convert inputs into outputs. It compares DMUs by calculating their efficiency scores based on a set of inputs and outputs. The method has been widely applied in the fields of agriculture, education, energy, finance, government, healthcare, manufacturing, retail, sport, and transportation \citep{Liu2013}.

In DEA research, it is common to follow the efficiency measurement with a second-stage regression analysis that uses efficiency scores as the dependent variable and includes contextual (or environmental) variables as independent variables. This approach is known as two-stage DEA. In many cases, efficiency is assessed annually, which may require a panel regression as the second-stage model to account for time-varying contextual variables. The most frequently employed panel methods for the second stage are panel linear regression (see, e.g., \citealp{Chen2019a, Mamatzakis2013}) and panel Tobit regression (see, e.g., \citealp{Borozan2018, Fonchamnyo2016}). In linear regression, log transformations of efficiency scores are often used (see, e.g., \citealp{Poveda2011, Zhang2018}). Other panel methods include panel quantile regression (see, e.g., \citealp{Fryd2021, Zhang2018}), panel fractional regression (see, e.g., \citealp{DaSilvaESouza2015, Fonchamnyo2016}), and panel beta regression (see, e.g., \citealp{Pirani2018, Song2016}). \cite{Liu2016a} identifies two-stage DEA as one of the active fronts in DEA research.

The standard two-stage DEA has been subject to criticism by \cite{Simar2007}, \cite{Simar2011}, and \cite{Kneip2015}. The criticisms mainly stem from three issues: (1) correlation among the estimated efficiency scores due to the complex structure of the data generating process, (2) the use of estimated efficiency scores as dependent variable instead of the true unobserved efficiency scores, and (3) the potential inseparability between the frontier production and the impact of contextual variables. These issues can significantly affect the validity of inference. When dealing with repeated assessement of efficiency, there is also the issue of temporal dependence. Nevertheless, some authors such as \cite{Banker2008}, \cite{McDonald2009}, and \cite{Banker2019} argue for the use of linear regression. For a survey on statistical approaches in nonparametric frontier models, see \cite{Moradi-Motlagh2022}.

The focus of this paper is on the two-stage DEA in the multiple-period setting. However, the time series dimension can be incorporated into DEA models in various ways. Notable examples include window analysis \citep{Charnes1985}, Malmquist productivity index \citep{Fare1994}, dynamic DEA model \citep{Nemoto1999, Sueyoshi2005}, multiperiod aggregative efficiency \citep{Park2009}, and dynamic slacks-based measure \citep{Tone2010}.

In this paper, we present an alternative approach for the panel second stage of DEA. Instead of modeling efficiency scores, we propose to model the rankings. In the recent literature, \cite{Holy2022f} developed a time series model for rankings that utilize the Plackett--Luce distribution and incorporates autoregressive and score dynamics. This model is based on the modern framework of score-driven models introduced by \cite{Creal2013} and \cite{Harvey2013}. While \cite{Holy2022f} applied the model to the results of the Ice Hockey World Championships, they also suggested its potential use in the second stage of DEA. Following this call, we devote this paper to exploring the use of this dynamic ranking model in DEA.

The motivation for using the score-driven dynamic ranking model in the second stage of DEA arises from the following properties:
\begin{itemize}
\item \emph{Relevance of Rankings.} Rankings preserve the important information of mutual comparison among DMUs. In certain scenarios, the primary objective of DEA may even be to obtain rankings of DMUs, in which case modeling rankings directly is more appropriate. The long-term behavior of DMUs may also be of interest, in which case the long-term ranking may have a clearer interpretation than an aggregate of efficiency scores.
\item \emph{Robustness to DEA Model.} Consider two DEA models: the super-efficiency DEA model of \cite{Andersen1993} and the universal DEA model of \cite{Hladik2019}, both with either constant returns to scale (CRS) of \cite{Charnes1978} or variable returns to scale (VRS) of \cite{Banker1984}. Despite producing different efficiency scores, these models generate the exact same ranking. By modeling rankings instead of efficiency scores in the second stage, any differences between these models are eliminated, as the response variable is the same in both cases. An additional consideration when modeling efficiency scores is whether to use the logarithmic transformation. However, since the log transformation preserves rankings, this is not a concern when using a ranking model.
\item \emph{Robustness to Outliers.} Outliers, in the form of extreme values of efficiency scores, can significantly influence the coefficients in a second-stage regression model. However, using rankings can mitigate this issue, as a DMU with an extremely low or high efficiency score would simply be ranked last or first, respectively. Thus, a ranking model can effectively handle such outliers.
\item \emph{Simple yet Powerful.} The model of \cite{Holy2022f} is straightforward to work with. The Plackett--Luce distribution, unlike its alternatives, is available in a closed form (see \citealp{Alvo2014}) and the dynamics are observation-driven (see \citealp{Cox1981}). As a result, the model can be estimated using the maximum likelihood method, and conventional Hessian-based standard errors can be used.  Moreover, the model only requires a modest number of parameters, consisting of individual effects of DMUs, regression coefficients common for all DMUs, and two additional parameters controling dynamics common for all DMUs.
\end{itemize}
Our approach also faces the following limitations:
\begin{itemize}
\item \emph{Loss of Information.} While using rankings instead of efficiency scores can provide robustness to DEA model and outliers (as discussed above), it also leads to loss of information. This loss can be beneficial in some scenarios, but it is still important to recognize that it occurs. One drawback of using rankings alone is that it is not possible to determine the boundary between inefficient and efficient DMUs. Efficiency scores, on the hand, provide a clear distinction between the two groups.
\item \emph{Different Data Generating Process.} Our approach does not address the criticism of \cite{Simar2007}, \cite{Simar2011}, and \cite{Kneip2015}. Indeed, the dependence between the DMUs is not captured by the Plackett--Luce distribution, which assumes the property known as the independence of irrelevant alternatives. The data generating process assumed by the model of \cite{Holy2022f} is much simpler then the true one generated by DEA.
\item \emph{Absence of Ties.} The model of \cite{Holy2022f} has a limitation in that it does not allow for rankings with ties. This means that in the second stage, we need to use a suitable DEA model that can rank all DMUs, including the efficient ones. However, this can be addressed by extending the Plackett-Luce distribution to incorporate ties, as demonstrated by \cite{Turner2020}.
\item \emph{Sufficient Variation in Rankings.} A single realization of efficiency scores is often used in a second stage regression model. A single ranking is, however, not enough for a meaningful analysis. Repeated rankings are therefore needed, which naturally take the form of panel data. Our approach is therefore suitable only when the time dimension is present. Even with repeated rankings, however, the Plackett--Luce distribution requires that for any possible partition of DMUs into two non-empty subsets, there exists at least one DMU in the second subset that is ranked higher than at least one DMU in the first subset (see \citealp{Hunter2004}).
\end{itemize}
Our approach is fundamentally different from traditional panel regressions, but it is not intended to replace them. Particularly when it suffers from the same shortcomings highlighted by \cite{Simar2007}, \cite{Simar2011}, and \cite{Kneip2015}. Instead, our approach is best used as a complement to traditional panel regressions to provide valuable insights that are not burdened by the problems specific to efficiency scores. This can be viewed as a form of robustness check, where both approaches are used to provide a more complete picture of the data. Given the controversies surrounding the second stage DEA, conducting extensive robustness checks is crucial for ensuring the reliability and validity of the results. DEA practitioners who wish to utilize the dynamic ranking model can do so easily using the \verb"gasmodel" R package, which offers all the necessary tools for estimation, forecasting, and simulation.

As an illustration of the proposed approach, we explore the research efficiency in higher education of European Union (EU) countries through the analysis of scientific publications in 2005--2020. In the first stage, we perform DEA analysis for each year independently. We use gross domestic expenditure on R\&D and the number of researchers as inputs to reflect the financial and human resources, respectively. For outputs, we use the number of publications and the number of citations to reflect the quantity and quality of scientific research, respectively. In the second stage, we investigate the influence of good governance on the research efficiency. As contextual variables, we use the six Worldwide Governance Indicators (WGI) of \cite{Kaufmann2011}, together with the gross domestic product (GDP). We perform panel linear regression analysis of efficiency scores obtained by three DEA models proposed by \cite{Charnes1978}, \cite{Andersen1993}, and \cite{Hladik2019}, along with the dynamic ranking model of \cite{Holy2022f}. As a side note, we demonstrate that the results obtained from the model of \cite{Andersen1993} can be derived from the model of \cite{Hladik2019}. All models uncover that the Voice and Accountability indicator is significantly positively correlated with research efficiency suggesting that participation in selecting the government, freedom of expression, freedom of association, and freedom of media are key factors of governance influencing research efficiency. The Government Effectiveness indicator has also positive effect, however, its significance is not confirmed by all models and this result is therefore not robust. No other significant relations are found. By utilizing the proposed approach in this study, we are able to assess the robustness of the relationship to the Voice and Accountability indicator. However, the results also indicate caution in interpreting the findings related to the Government Effectiveness indicator. Therefore, conducting extensive robustness checks such as this one is important to increase the reliability of the analysis and prevent misleading conclusions.

The rest of the paper is structured as follows. In Section \ref{sec:first}, we present three DEA models proposed by \cite{Charnes1978}, \cite{Andersen1993}, and, \cite{Hladik2019}, which are utilized in the subsequent analysis. In Section \ref{sec:second}, we present details on the dynamic ranking model of \cite{Holy2022f} and its estimation, along with some modifications suitable to our case. In Section \ref{sec:emp}, we conduct an empirical study to examine research efficiency in higher education and compare the proposed ranking approach with the traditional panel regression approach. We conclude the paper in Section \ref{sec:con}.

\section{First Stage: Measuring Efficiency}
\label{sec:first}

The first stage of DEA involves determining the relative efficiency scores of the DMUs. The number of DMUs is denoted by $N$. Each DMU transforms $I$ inputs into $J$ outputs. Let $x_{ni}$ denote the $i$-th input of the $n$-th DMU, and $y_{nj}$ denote the $j$-th output of the $n$-th DMU. The matrix of inputs is denoted by $X = (x_{ni})_{n=1,i=1}^{N,I}$, while the matrix of outputs is denoted by $Y = (y_{nj})_{n=1,j=1}^{N,J}$. The inputs of a single DMU $n$ are denoted by $x_{n} = (x_{n1}, \ldots, x_{nI})^{\intercal}$, and the outputs of a DMU $n$ are denoted by $y_{n} = (y_{n1}, \ldots, y_{nJ})^{\intercal}$. The notation $X_{-n}$ represents the inputs of every DMU but $n$, while $Y_{-n}$ represents the outputs of every DMU but $n$.

\subsection{Basic DEA}
\label{sec:charnes}

\cite{Charnes1978} proposed the very first DEA model, which has since become one of the most widely used DEA models to date. This model is commonly referred to as the CCR model and is based on the assumption of constant returns to scale (CRS). The efficiency scores $\theta^{CCR}_n$ are found for each DMU $n$ by the following linear program:
\begin{equation}
\label{eq:charnes}
\begin{aligned}
\theta^{CCR}_n & = \max_{\substack{u, v}} \ y_{n}^{\intercal} u \\
& \phantom{==} \text{subject to} & x_{n}^{\intercal} v & = 1, \\
&& Y u - X v & \leq 0, \\
&& u & \geq 0, \\
&& v & \geq 0, \\
\end{aligned}
\end{equation}
where $u$ and $v$ are vectors of weights for the outputs and inputs respectively. The efficiency scores for inefficient DMUs lie in $[0,1)$ and are equal to 1 for efficient DMUs. Note that DMUs achieving an efficiency score of 1 may still be considered weakly efficient, as they may have non-zero slack in either inputs or outputs (see, e.g., \citealp{Cooper2007}).

\subsection{Super-Efficiency DEA}
\label{sec:andersen}

A shortcoming of the CCR model is that it cannot differentiate between efficient DMUs, which can lead to the loss of valuable information. \cite{Andersen1993} proposed a super-efficiency DEA to overcome this limitation. In this model, the DMU under evaluation is excluded from the set of benchmarks, which allows efficient DMUs to achieve score greater than 1. The super-efficiency model with CRS (labeled as the AP model) is given by the following linear program:
\begin{equation}
\label{eq:andersen}
\begin{aligned}
\theta^{AP}_n & = \max_{\substack{u, v}} \ y_{n}^{\intercal} u \\
& \phantom{==} \text{subject to} & x_{n}^{\intercal} v & = 1, \\
&& Y_{-n} u - X_{-n} v & \leq 0, \\
&& u & \geq 0, \\
&& v & \geq 0. \\
\end{aligned}
\end{equation}
The efficiency scores for inefficient DMUs are the same as those obtained from the CCR model, while the scores for efficient DMUs are greater than or equal to 1.

\subsection{Universal DEA}
\label{sec:hladik}

Recently, \cite{Hladik2019} proposed a DEA formulation that focuses on a robust optimization viewpoint. The model uses a scaled Chebyshev norm to measure efficiency as a distance to inefficiency and inefficiency as a distance to efficiency. The scores generated by this model are universal in the sense that they are naturally normalized, and therefore, can be compared across unrelated models. The universal DEA model with CRS (labeled as the H model) is given by the following linear program:
\begin{equation}
\label{eq:hladik}
\begin{aligned}
\theta^H_n & = \max_{\substack{\delta, u, v}} \ 1 + \delta \\
& \phantom{==} \text{subject to} & y_{n}^{\intercal} u & \geq 1 + \delta, \\
&& x_{n}^{\intercal} v & \leq 1 - \delta, \\
&& Y_{-n} u - X_{-n} v & \leq 0, \\
&& u & \geq 0, \\
&& v & \geq 0. \\
\end{aligned}
\end{equation}
Note that \cite{Hladik2019} also proposed a nonlinear DEA model based on the Chebyshev norm, to which \eqref{eq:hladik} is a tight approximation. The efficiency scores for inefficient DMUs lie in $[0, 1)$, while the scores for efficient DMUs lie in $[1, 2]$.

Applications of the universal DEA model include \cite{Holy2018e}, which analyzes the efficiency of basic and applied research; \cite{Fryd2021}, which focus on the efficiency of farms; and \cite{Holy2022a}, which examines the efficiency of public libraries.

\subsection{Relation of Super-Efficiency and Universal DEA}

The universal DEA model is closely related to the super-efficiency DEA model of \cite{Andersen1993}. \cite{Hladik2019} showed that the ranking of DMUs according to $\theta^{AP}_n$ is the same as the ranking according to $\theta^H_n$. We go a bit further and show that the models are even more connected as the efficient scores themselves can be derived by the following transformations:
\begin{equation}
\label{eq:transform}
\theta_n^H = \frac{2 \theta_n^{AP}}{1 + \theta_n^{AP}}, \qquad \theta_n^{AP} = \frac{\theta_n^H}{2 - \theta_n^H}.
\end{equation}

The proof is as follows. First, we rewrite model \eqref{eq:andersen} as
\begin{equation}
\label{eq:andersen2}
\begin{aligned}
\theta^{AP}_n & = \max_{\substack{\alpha, u, v}} \ \alpha  \\
& \phantom{==} \text{subject to} & y_{n}^{\intercal} u & \geq \alpha, \\
&& x_{n}^{\intercal} v & \leq 1, \\
&& Y_{-n} u - X_{-n} v & \leq 0, \\
&& u & \geq 0, \\
&& v & \geq 0, \\
&& \alpha & \geq 0. \\
\end{aligned}
\end{equation}
The objective function $y_{n}^{\intercal} u$ in \eqref{eq:andersen} can be replaced by an additional variable $\alpha \geq 0$, when the constraint $y_{n}^{\intercal} u \geq \alpha$ is further incorporated. The constraint $x_{n}^{\intercal} v = 1$ in \eqref{eq:andersen} can be replaced by $x_{n}^{\intercal} v \leq 1$, as any solution with $c = x_{n}^{\intercal} v < 1$ must be suboptimal. This is because both $v$ and $u$ could be multiplied by $1/c$ to achieve $x_{n}^{\intercal} v = 1$, resulting in an objective that is $1/c$ times higher.

Next, using the substitution 
\begin{equation}
\label{eq:tilde}
\tilde{u} = \frac{u}{1 - \delta}, \qquad \tilde{v} = \frac{v}{1 - \delta}, \qquad \tilde{\alpha} = \frac{1 + \delta}{1 - \delta},
\end{equation}
we rewrite model \eqref{eq:hladik} as
\begin{equation}
\label{eq:hladik2}
\begin{aligned}
\theta^H_n & = \max_{\substack{\tilde{\alpha}, \tilde{u}, \tilde{v}}} \ \frac{2 \tilde{\alpha}}{1 + \tilde{\alpha}} \\
& \phantom{==} \text{subject to} & y_{n}^{\intercal} \tilde{u} & \geq \tilde{\alpha}, \\
&& x_{n}^{\intercal} \tilde{v} & \leq 1, \\
&& Y_{-n} \tilde{u} - X_{-n} \tilde{v} & \leq 0, \\
&& \tilde{u} & \geq 0, \\
&& \tilde{v} & \geq 0, \\
&& \tilde{\alpha} & \geq 0. \\
\end{aligned}
\end{equation}
Note that we can impose $\tilde{\alpha} \geq 0$ as $y_{n}^{\intercal} \tilde{u} \geq 0$. This also follows from $\delta \in [-1, 1)$ in model \eqref{eq:hladik}.

Models \eqref{eq:andersen2} and \eqref{eq:hladik2} have the same constraints and differ only in the objective function. The function $2 \tilde{\alpha} / (1 + \tilde{\alpha})$ is monotonically increasing on $[0, \infty)$, thereby attaining its maximum value at the same point as $\tilde{\alpha}$. Thus, we have $\theta_n^H =2 \theta_n^{AP} / \left( 1 + \theta_n^{AP} \right)$. This proof follows and extends Proposition 2 in \cite{Hladik2019}.

\section{Second Stage: Modeling Dynamic Rankings}
\label{sec:second}

The second stage of DEA involves identifying the factors that affect efficiency scores and measure their impact. We assume periodic evaluation of efficiency of the same set of DMUs at times $t = 1, \ldots, T$ with efficiency scores $\theta_t = \left( \theta_{1t}, \ldots, \theta_{Nt} \right)^{\intercal}$. In this paper, we propose to model rankings of DMUs, instead of their efficiency scores as is usual in the second-stage DEA. Let $R_{t}(n)$ denote the rank of a DMU $n$ according to efficiency scores $\theta_t$ at time $t$. The complete ranking at time $t$ is then denoted by $R_t = \left( R_t(1), \ldots, R_{t}(N) \right)^{\intercal}$. The inverse of this ranking is the ordering $O_t = \left( O_{t}(1), \ldots, O_{t}(N) \right)^{\intercal}$ at time $t$, where $O_{t}(r)$ represents the DMU with rank $r$ at time $t$. We employ the dynamic ranking model of \cite{Holy2022f}.

\subsection{Plackett--Luce Distribution}
\label{sec:pluce}

We assume that at each time $t$ the ranking $R_t$ follows the Plackett--Luce distribution proposed by \cite{Luce1959} and \cite{Plackett1975}. In the ranking literature, it is a widely used probability distribution for random variables in the form of permutations. Each DMU $n$ at each time $t$ has a worth parameter $w_{nt} \in \mathbb{R}$ reflecting its rank at time $t$. The probability of a higher rank increases with a higher worth parameter value. Specifically, the probability mass function is given by
\begin{equation}
\label{eq:prob}
f \left( R_t \middle| w_t \right) = \prod_{r=1}^{N} \frac{\exp \left( w_{O_t(r)t} \right)}{ \sum_{s=r}^N \exp \left( w_{O_t(s)t} \right) }.
\end{equation}
In other words, a ranking is iteratively constructed by selecting the best DMU, followed by the second best, the third best, and so on. At each stage, the probability of selecting a particular DMU is proportional to the exponential of its worth parameter divided by the sum of the exponentials of the worth parameters of all DMUs that have not been selected yet. The log-likelihood function is given by
\begin{equation}
\label{eq:loglik}
\ell \left( w_t \middle| R_t \right) = \sum_{n=1}^{N} w_{nt} - \sum_{r=1}^{N} \ln \left( \sum_{s=r}^N \exp w_{O_t(s)t} \right).
\end{equation}
The score (i.e.\ the gradient of the log-likelihood function) is given by
\begin{equation}
\label{eq:score}
\nabla_n \left( w_t \middle| R_t \right) = 1 - \sum_{r=1}^{R_t(n)} \frac{\exp \left( w_{nt} \right)}{\sum_{s=r}^N \exp \left( w_{O_t(s)t} \right)}, \qquad n = 1, \ldots, N.
\end{equation}

The Plackett--Luce distribution is based on the Luce's choice axiom, which states that the probability of selecting one item over another from a set of items is not influenced by the presence or absence of other items in the set (see \citealp{Luce1977}). This property of choice is known as the independence of irrelevant alternatives. Clearly, this property is not met in the case of DEA as addition or removal of DMUs from the set can influence efficiency scores and even ranking of other DMUs. As in the case of many second-stage models, the proposed dynamic ranking model therefore does not conform to the complex data generating process of DEA efficiency scores and rankings. Nevertheless, the proposed model can be a useful tool due to its simplicity when applied with caution.

\subsection{Regression and Dynamics}
\label{sec:reg}

We let the worth parameters linearly depend on $K$ contextual variables and also include an autoregressive and score-driven component. For $n = 1,\ldots, N$, $t = 1, \ldots, T$, the worth parameters are then given by the recursion
\begin{equation}
\label{eq:dyn}
w_{nt} = \omega_n + \sum_{k=1}^K \beta_k z_{nkt} + e_{nt}, \quad e_{nt} = \varphi e_{nt-1} + \alpha \nabla_n \left( w_{t-1} \middle| R_{t-1} \right),
\end{equation}
where $\omega_n$ are the individual effects for each DMU $n$, $\beta_k$ are the regression parameters for the contextual variables $z_{nkt}$, $\varphi$ is the autoregressive parameter, and $\alpha$ is the score parameter for the lagged score $\nabla_n \left( w_{t-1} \middle| R_{t-1} \right)$ given by \eqref{eq:score}. The model corresponds to panel regression with fixed effects and dynamic error term. Note that the model is overparametrized as the probability mass function \eqref{eq:prob} is invariant to the addition of a constant to all worth parameters. We therefore use standardization
\begin{equation}
\label{eq:stan}
\sum_{n=1}^N \omega_n = 0.
\end{equation}

Our specification differs from the model of \cite{Holy2022f} by introducing the separate $e_{nt}$ component. Our specification is inspired by the regression with ARMA errors, while the specification of \cite{Holy2022f} resemble the ARMAX model. In our specification, the contextual variables influence only concurrent ranking, which is easier to interpret. Our model is also easier for numerical estimation as $\omega_n$ and $\varphi$ are disconnected.

The $e_{nt}$ component captures dynamic effects by the autoregressive term and the lagged score. The model therefore belongs to the class of score-driven models, also known as generalized autoregressive score (GAS) models or dynamic conditional score (DCS) models, proposed by \cite{Creal2013} and \cite{Harvey2013}. The score can be interpreted as a measure of the fit of the Plackett--Luce model to the observed rankings. A positive score indicates that a DMU $n$ is ranked higher than what its worth parameter $w_{nt}$ suggests, while a negative score suggests that it is ranked lower. A score of zero indicates that the DMU is ranked as expected according to its worth parameter. Thus, the score can be used as a correction term for the worth parameter after the ranking is observed.

\subsection{Maximum Likelihood Estimation}
\label{sec:mle}

The model is observation-driven and can be estimated by the maximum likelihood method. Let $\theta = (\omega_1, \ldots, \omega_{N-1}, \beta_1, \ldots, \beta_K, \varphi, \alpha)'$ denote the vector of the $N+K+1$ parameters to be estimated. Note that $\omega_N$ is obtained from \eqref{eq:stan} as $\omega_N = - \sum_{n=1}^{N-1}\omega_n$. The maximum likelihood estimate $\hat{\theta}$ is then given by
\begin{equation}
\label{eq:mle}
\hat{\theta} \in \arg\max_{\theta} \sum_{t=1}^T \ell \left( w_t \middle| R_t \right),
\end{equation}
where the log-likelihood $\ell \left( w_t \middle| R_t \right)$ is given by \eqref{eq:loglik} and $w_t$ follow \eqref{eq:dyn}. The problem \eqref{eq:mle} can be numerically solved by any general-purpose algorithm for nonlinear optimization. Furthermore, the standard errors of the estimated parameters are computed using the empirical Hessian of the log-likelihood evaluated at $\hat{\theta}$.

In order for the log-likelihood to have a unique maximum, it is necessary that for any possible partition of DMUs into two non-empty subsets, there exists at least one DMU in the second subset that is ranked higher than at least one DMU in the first subset (see \citealp{Hunter2004}). This condition ensures that no DMU is always ranked first, which would result in an infinite worth parameter and violate the assumptions of maximum likelihood estimation.

\begin{landscape}
\begin{table}
\begin{center}
\caption{An overview of relevant studies.}
\label{tab:lit}
\begin{tabular}{>{\raggedright}m{5.1cm}>{\raggedright}m{4cm}>{\raggedright}m{7cm}>{\raggedright}m{7cm}}
\toprule
Paper & Sample & Inputs & Outputs \tabularnewline
\midrule
\cite{Aristovnik2012} & 37 countries & R\&D expenditure, Researchers & Articles, Patent apps., High-tech exports \tabularnewline \tabularnewline
\cite{Belgin2019} & 12 Turkish regions & R\&D expenditure, R\&D personnel & Patents granted, High-tech exports \tabularnewline \tabularnewline
\cite{Carracedo2022} & 75 countries & R\&D expenditure & Articles, Patent apps., Trademark apps. \tabularnewline \tabularnewline
\cite{Chen2011} & 24 countries & R\&D expenditure stocks, R\&D personnel & Articles, Patent apps., Royalty fees \tabularnewline \tabularnewline
\cite{Cullmann2012} & 28 countries & Detailed R\&D expenditure, Researchers & Weighted and unweighted patents \tabularnewline \tabularnewline
\cite{Ekinci2017} & 28 EU countries & Detailed R\&D expenditure, R\&D personnel, Employment & Patents granted, Publications \tabularnewline \tabularnewline
\cite{Halaskova2020} & 28 EU countries & R\&D expenditure, Budget appropriations, Researchers & Patents granted, High-tech exports, Publications \tabularnewline \tabularnewline
\cite{Han2016} & 15 Korean regions & R\&D expenditure & Patent apps., Publications \tabularnewline \tabularnewline
\cite{Hung2009} & 27 countries & R\&D expenditure, Researchers & Article share, Citation share \tabularnewline \tabularnewline
\cite{Holy2018e} & 28 EU countries & R\&D expenditure, Researchers & Citations, Patent apps. \tabularnewline \tabularnewline
\cite{Lee2005} & 27 countries & R\&D expenditure, Researchers & Patents, Articles, Tech balance of receipts \tabularnewline \tabularnewline
\cite{Roman2010} & 14 BG and RO regions & R\&D expenditure, R\&D personnel & Patent apps. \tabularnewline \tabularnewline
\cite{Sharma2008} & 22 countries & R\&D expenditure, Researchers & Patents granted, Publications \tabularnewline \tabularnewline
\cite{Thomas2011} & 50 US states and D.C. & R\&D expenditure & Patents granted, Publications \tabularnewline \tabularnewline
\cite{Zuo2017} & 30 Chinese regions & R\&D expenditure, Researchers & Patent apps., High-tech exports, Publications \tabularnewline
\bottomrule
\end{tabular}
\end{center}
\end{table}
\end{landscape}

\section{Empirical Study}
\label{sec:emp}

Our empirical study aims to analyze research efficiency in the higher education sector by examining scientific publications on a country-level basis, with a particular focus on the EU countries between 2005 and 2020. Specifically, we seek to determine whether certain aspects of good governance have a positive impact on research efficiency.

\subsection{Relevant Studies}
\label{sec:lit}

Assessing the efficiency of research and development (R\&D) is a widely studied topic in the data envelopment analysis (DEA) literature. In Table \ref{tab:lit}, we present a list of several relevant DEA papers and the key specifics of each study. We focus on the assessment of countries (and regions), although similar analyses can be performed at more detailed levels of institutions (see, e.g., \citealp{Jablonsky2016}) and projects (see, e.g., \citealp{Lee2009}). Typically, studies on R\&D efficiency use financial resources and human resources as the two main inputs. In terms of outputs, some studies focus on variables related to scientific publications (such as \citealp{Hung2009}), some on patents  (such as \citealp{Cullmann2012}), while the majority consider both types of R\&D-related outcomes.

\subsection{Input, Output, and Contextual Variables}
\label{sec:var}

As inputs, we use the following variables:
\begin{itemize}
\item \emph{R\&D Expenditure} refers to the gross domestic expenditure to R\&D activities performed in the higher education sector. The unit is million purchasing power standards. \cite{Holy2018e} emphasize the importance of accounting for purchasing power parity when adjusting prices to ensure meaningful comparisons between countries with varying purchasing power. This variable reflects the financial resources.
\item \emph{Number of Researchers} refers to the total number of researchers employed in the higher education sector. The unit is full-time equivalent. This variable reflects the human resources.
\end{itemize}
As outputs, we use the following variables:
\begin{itemize}
\item \emph{Number of Publications} represents the number of articles, reviews, and conference papers published. This variable reflects the quantity of scientific research.
\item \emph{Number of Citations} represents the number of citations to the published articles, reviews, and conference papers. This variable reflects the quality of scientific research.
\end{itemize}
As contextual variables, we use the six Worldwide Governance Indicators (WGI), which \cite{Kaufmann2011} define in the following way:
\begin{itemize}
\item \emph{Voice and Accountability} captures perceptions of the extent to which a country's citizens are able to participate in selecting their government, as well as freedom of expression, freedom of association, and a free media.
\item \emph{Political Stability and Absence of Violence/Terrorism} captures perceptions of the likelihood that the government will be destabilized or overthrown by unconstitutional or violent means, including politically‐motivated violence and terrorism.
\item \emph{Government Effectiveness} captures perceptions of the quality of public services, the quality of the civil service and the degree of its independence from political pressures, the quality of policy formulation and implementation, and the credibility of the government's commitment to such policies.
\item \emph{Regulatory Quality} captures perceptions of the ability of the government to formulate and implement sound policies and regulations that permit and promote private sector development.
\item \emph{Rule of Law} captures perceptions of the extent to which agents have confidence in and abide by the rules of society, and in particular the quality of contract enforcement, property rights, the police, and the courts, as well as the likelihood of crime and violence.
\item \emph{Control of Corruption} captures perceptions of the extent to which public power is exercised for private gain, including both petty and grand forms of corruption, as well as ``capture'' of the state by elites and private interests.
\end{itemize}
Finally, we also include the following variable as a contextual variable:
\begin{itemize}
\item \emph{Gross Domestic Product} is used to control for the economic development of a country. To filter out the trend, we use the percentage of EU total GDP per capita based on million purchasing power standards.
\end{itemize}
We therefore have $I=2$ input variables, $J=2$ output variables, and $K=7$ contextual variables. Similarly to \cite{Holy2018e}, we lag the input and contextual variables by one year, recognizing that there is typically a delay between the input variables and the corresponding output variables.

\subsection{Data Sample}
\label{sec:data}

Our data sample contains all $N=27$ countries of EU. The outputs are taken from 2005 to 2020, while the inputs and contextual variables are taken with a one-year lag from 2004 to 2019. We therefore have $T=16$ time periods to analyze. The source of the R\&D expenditure, the number of researchers, and the GDP is Eurostat\footnote{\url{https://ec.europa.eu/eurostat/data/database}}. There were 4 missing observations for the number of researchers of Greece in 2004, 2008, 2009, and 2010. We have interpolated these values using linear regression. The source of the number of documents and the number of citations is Scimago Journal \& Country Rank\footnote{\url{https://www.scimagojr.com}}. The source of the Worldwide Governance Indicators is the World Bank\footnote{\url{https://info.worldbank.org/governance/wgi}}.

\begin{figure}
\begin{center}
\includegraphics[width=15cm]{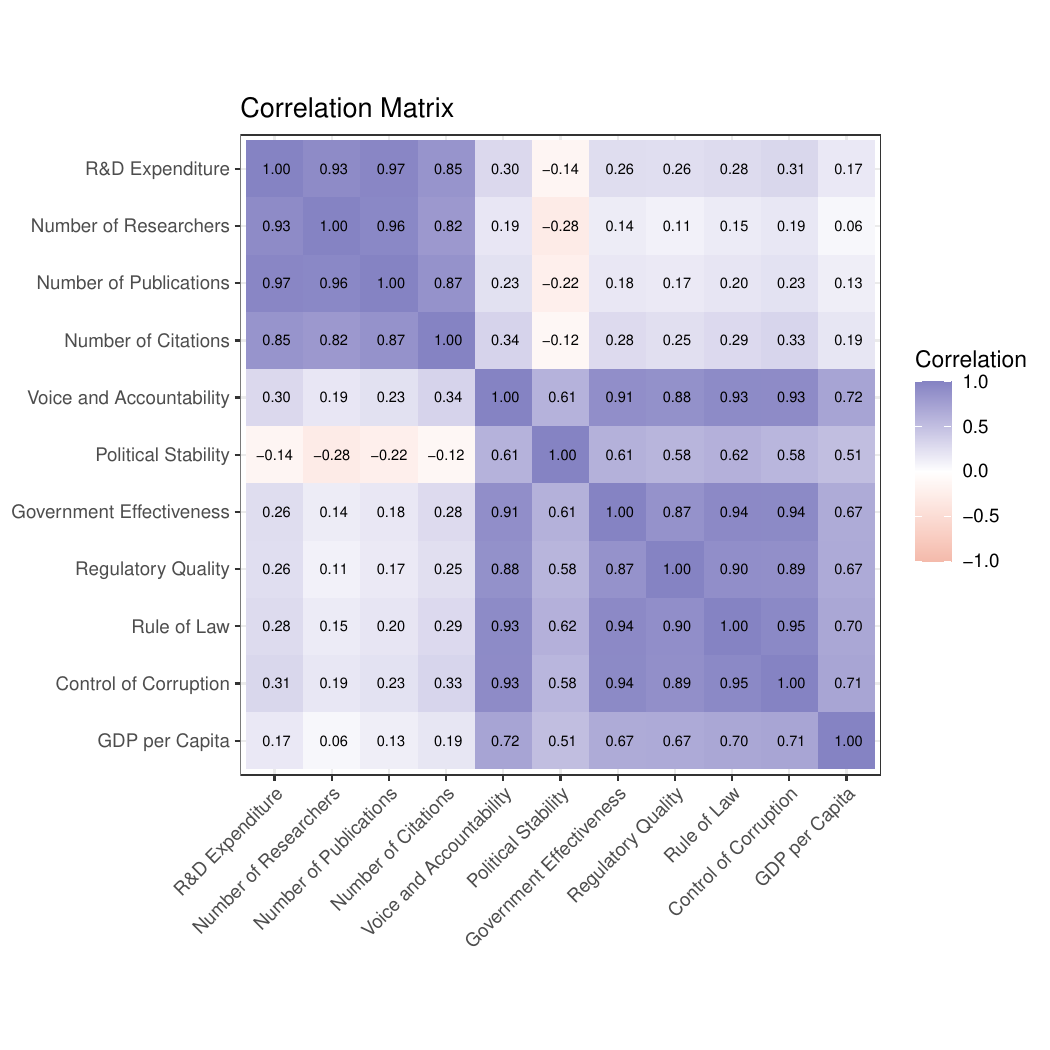}
\caption{The correlation matrix of the input, output, and contextual variables.}
\label{fig:cor}
\end{center}
\end{figure}

Table \ref{tab:data} presents the five-number summary of all variables. Figure \ref{fig:cor} depicts the correlation matrix. It reveals strong positive correlations among all input and output variables. The contextual variables form a separate block, in which all variables also demonstrate positive correlations, albeit to a lesser degree in some cases. The correlations between input/output variables and contextual variables are relatively small.

\begin{table}
\begin{center}
\caption{The minimum (Min), first quartile (Q1), median (Q2), third quartile (Q3), and maximum (Max) of the input, output, and contextual variables.}
\label{tab:data}
\begin{tabular}{lrrrrr} 
\toprule
Variable & Min & Q1 & Q2 & Q3 & Max \\ 
\midrule
R\&D Expenditure & 4.72 & 133.71 & 628.35 & 2\,163.98 & 18\,035.70 \\ 
Number of Researchers & 143.00 & 3\,604.00 & 10\,339.00 & 20\,045.00 & 117\,638.00 \\ \\
Number of Publications & 99.00 & 3\,895.50 & 13\,563.00 & 29\,190.00 & 175\,034.00 \\ 
Number of Citations & 2\,036.00 & 64\,936.50 & 280\,954.00 & 761\,125.50 & 4\,689\,760.00 \\ \\
Voice and Accountability & 0.26 & 0.91 & 1.09 & 1.39 & 1.80 \\ 
Political Stability & -0.47 & 0.51 & 0.79 & 1.02 & 1.62 \\ 
Government Effectiveness & -0.37 & 0.67 & 1.05 & 1.57 & 2.35 \\ 
Regulatory Quality & 0.14 & 0.87 & 1.15 & 1.56 & 2.05 \\ 
Rule of Law & -0.19 & 0.59 & 1.06 & 1.71 & 2.12 \\ 
Control of Corruption & -0.38 & 0.31 & 0.91 & 1.66 & 2.46 \\ \\
GDP per Capita & 0.35 & 0.70 & 0.91 & 1.21 & 2.83 \\ 
\bottomrule
\end{tabular}
\end{center}
\end{table}

\subsection{Suitability of DEA Model}
\label{sec:suit}

In order for efficiency scores to be interpretable, several criteria need to be met. We have adopted the best practices in DEA as outlined by \cite{Dyson2001} and \cite{Cook2014}. We begin by establishing that the process under evaluation is well-defined. Our focus is on the research output in the form of scientific publications. The two chosen output variables encompass both the quantity and quality of scientific publications. While quantity is naturally quantifiable, measuring quality can be achieved using several metrics such as the number of citations and the h-index. However, combining indices and volume measures can pose difficulties and we have therefore decided to use the number of citations for our analysis. The two primary resources for conducting research are funding and personnel, both of which are represented by the two input variables we have selected. All input and output variables are volume measures and are isotonic (i.e.\ increased input reduces efficiency, while increased output increases efficiency). With a total of 4 input and output variables and 27 DMUs, our DEA model possesses sufficient discriminatory power.

Next, we examine the homogeneity assumption. Our set of DMUs encompasses all EU countries as of February 2020, although it should be noted that EU membership changed during the period under observation. Specifically, Romania and Bulgaria joined in 2007, and Croatia became a member in July 2013, whereas the United Kingdom departed in January 2020. Nevertheless, EU countries should be considered homogeneous in terms of research due to the harmonized policies and frameworks implemented by the European Commission, such as the European Research Area (ERA) and the Horizon Europe program. These initiatives aim to promote collaboration and standardization among EU member states, facilitating the dissemination of research findings and enhancing the overall quality of scientific output.

Finally, we analyze the appropriate returns to scale. Note that EU countries exhibit considerable variation in size, with Germany being the most populous country at 83.17 million people and Malta being the least populous with a population of 0.51 million as of January 2020. Our focus is on the higher education sector, which is composed of (1) universities, colleges of technology, and other institutions providing formal tertiary education programmes, (2) research institutes, centres, experimental stations and clinics that have their R\&D activities under the direct control of, or administered by, tertiary education institutions (see \citealp{OECD2015}). The scientific output of a country can be seen as the sum of outputs from these individual institutions. As a result, we assume that country size does not have a significant impact on the relative scientific output and employ the constant returns to scale (CRS) assumption in our analysis.

\subsection{Efficiency Scores}
\label{sec:eff}

\begin{figure}
\begin{center}
\includegraphics[width=15cm]{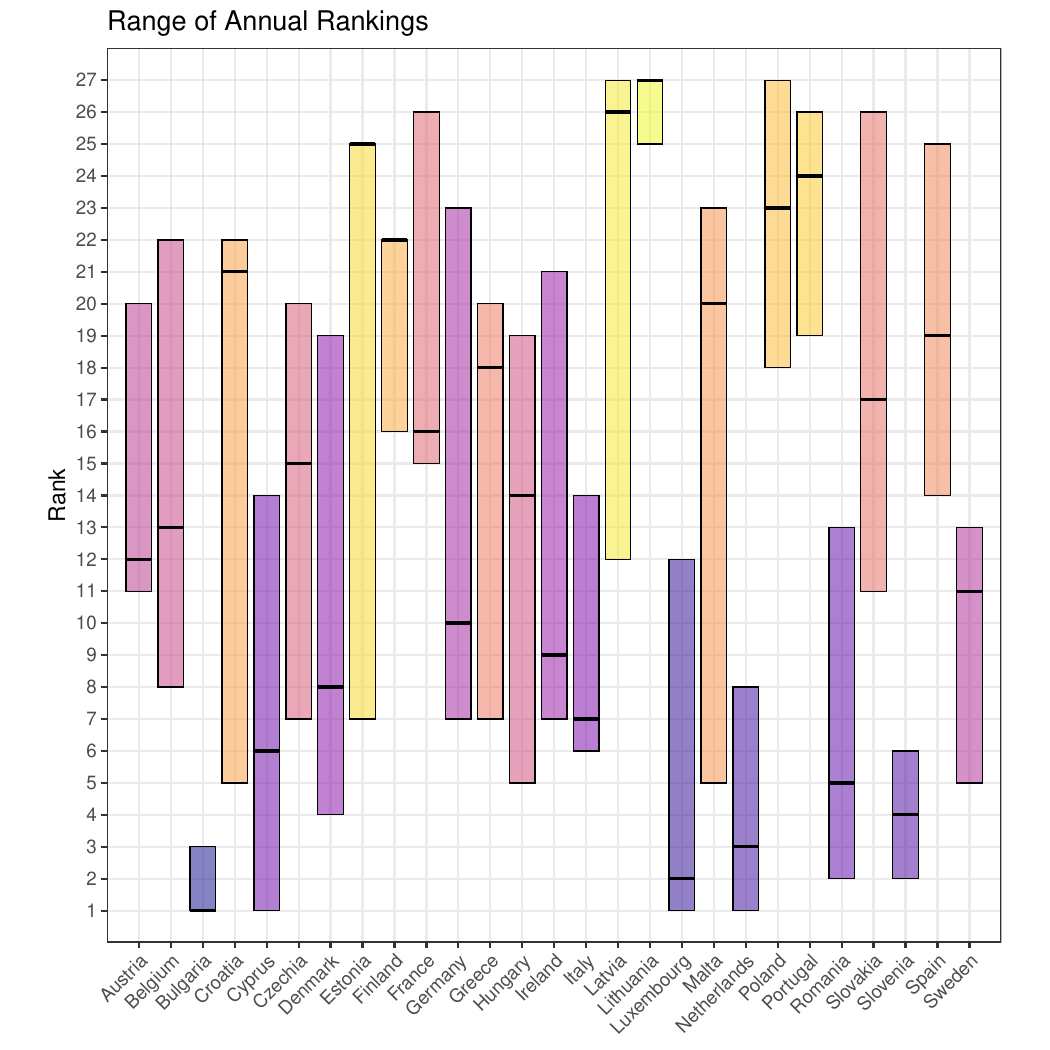}
\caption{The ranges of annual rankings by research efficiency, together with the long-term ranking according to the dynamic ranking model.}
\label{fig:interval}
\end{center}
\end{figure}

Table \ref{tab:eff} reports the efficiency scores obtained from the universal DEA model \eqref{eq:hladik} of \cite{Hladik2019} for all countries and years. Figure \ref{fig:interval} then displays the ranges of the annual rankings. Bulgaria consistently shows high levels of efficiency across most years, which can be primarily attributed to its extremely low R\&D spending, both in absolute value and relative to the number of publications, citations and even researchers. Romania is also found to be efficient in many years due to their relatively low R\&D spending. Cyprus has an average of 3.10 publications per researcher, the highest among all countries, followed by Slovenia with 2.57. Moving to Western Europe, the Netherlands stands out as the country with the highest number of citations per researcher, with an average of 81.49. The final country that is ever found efficient in our sample is Luxembourg. Germany, as the largest country, dominate in absolute values of all inputs and outputs; its efficiency is, however, average. At the other end of the efficiency spectrum, we find Latvia with 0.66 publications and 9.09 citations per researcher on average, and Lithuania with 0.61 publications and 8.82 citations per researcher on average.

\begin{landscape}
\begin{table}
\begin{center}
\caption{The annual efficiency scores obtained from the universal DEA model \eqref{eq:hladik} of \cite{Hladik2019}.}
\label{tab:eff}
\begin{tabular}{lrrrrrrrrrrrrrrrr} 
\toprule
Country & 2005 & 2006 & 2007 & 2008 & 2009 & 2010 & 2011 & 2012 & 2013 & 2014 & 2015 & 2016 & 2017 & 2018 & 2019 & 2020 \\ 
\midrule
  Austria & 0.85 & 0.79 & 0.81 & 0.68 & 0.70 & 0.76 & 0.78 & 0.75 & 0.72 & 0.70 & 0.66 & 0.59 & 0.55 & 0.60 & 0.58 & 0.54 \\ 
  Belgium & 0.87 & 0.81 & 0.87 & 0.75 & 0.83 & 0.78 & 0.87 & 0.76 & 0.76 & 0.72 & 0.62 & 0.54 & 0.55 & 0.57 & 0.59 & 0.50 \\ 
  Bulgaria & 1.21 & 1.15 & 1.47 & 1.40 & 1.40 & 1.21 & 1.17 & 1.29 & 1.28 & 1.22 & 0.99 & 1.21 & 1.19 & 1.30 & 1.22 & 1.09 \\ 
  Croatia & 0.56 & 0.62 & 0.74 & 0.74 & 0.76 & 0.83 & 0.99 & 0.98 & 0.93 & 0.89 & 0.84 & 0.86 & 0.66 & 0.74 & 0.65 & 0.62 \\ 
  Cyprus & 0.79 & 0.73 & 0.84 & 0.77 & 1.00 & 1.13 & 1.08 & 1.14 & 1.19 & 1.19 & 1.32 & 1.23 & 1.28 & 1.21 & 1.26 & 1.30 \\ 
  Czechia & 0.94 & 0.71 & 0.77 & 0.75 & 0.75 & 0.81 & 0.83 & 0.79 & 0.65 & 0.77 & 0.75 & 0.67 & 0.66 & 0.66 & 0.62 & 0.57 \\ 
  Denmark & 0.97 & 0.94 & 0.95 & 0.80 & 0.83 & 0.80 & 0.87 & 0.79 & 0.76 & 0.76 & 0.62 & 0.59 & 0.58 & 0.61 & 0.64 & 0.52 \\ 
  Estonia & 0.46 & 0.42 & 0.60 & 0.54 & 0.50 & 0.65 & 0.72 & 0.70 & 0.64 & 0.64 & 0.56 & 0.61 & 0.68 & 0.66 & 0.59 & 0.65 \\ 
  Finland & 0.64 & 0.61 & 0.70 & 0.59 & 0.68 & 0.67 & 0.71 & 0.66 & 0.65 & 0.67 & 0.60 & 0.55 & 0.52 & 0.52 & 0.54 & 0.48 \\ \\
  France & 0.71 & 0.66 & 0.69 & 0.58 & 0.66 & 0.68 & 0.71 & 0.68 & 0.63 & 0.64 & 0.60 & 0.51 & 0.45 & 0.46 & 0.44 & 0.42 \\ 
  Germany & 0.86 & 0.83 & 0.85 & 0.70 & 0.75 & 0.74 & 0.75 & 0.71 & 0.66 & 0.66 & 0.62 & 0.56 & 0.48 & 0.50 & 0.48 & 0.46 \\ 
  Greece & 0.62 & 0.62 & 0.70 & 0.64 & 0.80 & 0.84 & 0.94 & 0.86 & 0.84 & 0.76 & 0.65 & 0.56 & 0.58 & 0.58 & 0.62 & 0.56 \\ 
  Hungary & 0.69 & 0.65 & 0.69 & 0.73 & 0.71 & 0.72 & 0.78 & 0.78 & 0.73 & 0.78 & 0.75 & 0.76 & 0.80 & 0.70 & 0.63 & 0.57 \\ 
  Ireland & 0.85 & 0.83 & 0.89 & 0.77 & 0.83 & 0.83 & 0.95 & 0.86 & 0.78 & 0.72 & 0.58 & 0.58 & 0.55 & 0.56 & 0.58 & 0.54 \\ 
  Italy & 0.96 & 0.80 & 0.84 & 0.72 & 0.80 & 0.79 & 0.83 & 0.81 & 0.77 & 0.77 & 0.74 & 0.67 & 0.61 & 0.63 & 0.64 & 0.63 \\ 
  Latvia & 0.36 & 0.25 & 0.27 & 0.33 & 0.30 & 0.54 & 0.63 & 0.47 & 0.44 & 0.47 & 0.50 & 0.48 & 0.64 & 0.55 & 0.46 & 0.46 \\ 
  Lithuania & 0.32 & 0.31 & 0.35 & 0.45 & 0.40 & 0.48 & 0.48 & 0.41 & 0.39 & 0.38 & 0.36 & 0.33 & 0.41 & 0.45 & 0.43 & 0.44 \\ 
  Luxembourg & 1.26 & 1.31 & 1.13 & 1.22 & 1.07 & 0.89 & 0.92 & 0.87 & 0.86 & 0.85 & 0.72 & 0.82 & 0.76 & 0.77 & 0.71 & 0.57 \\ \\
  Malta & 0.53 & 0.62 & 0.61 & 0.59 & 0.62 & 0.74 & 0.70 & 0.82 & 0.71 & 0.72 & 0.70 & 0.90 & 0.76 & 0.89 & 0.76 & 0.83 \\ 
  Netherlands & 1.09 & 1.08 & 1.13 & 0.97 & 1.04 & 1.08 & 1.17 & 1.05 & 1.03 & 0.97 & 0.81 & 0.77 & 0.77 & 0.78 & 0.82 & 0.67 \\ 
  Poland & 0.57 & 0.60 & 0.62 & 0.66 & 0.62 & 0.63 & 0.59 & 0.61 & 0.56 & 0.60 & 0.55 & 0.55 & 0.47 & 0.44 & 0.39 & 0.36 \\ 
  Portugal & 0.52 & 0.51 & 0.63 & 0.57 & 0.46 & 0.50 & 0.60 & 0.62 & 0.66 & 0.55 & 0.48 & 0.48 & 0.44 & 0.46 & 0.49 & 0.47 \\ 
  Romania & 0.80 & 0.80 & 0.80 & 0.82 & 0.77 & 1.00 & 0.96 & 0.99 & 1.05 & 1.10 & 1.21 & 0.96 & 1.11 & 1.03 & 1.09 & 1.06 \\ 
  Slovakia & 0.65 & 0.72 & 0.63 & 0.70 & 0.64 & 0.80 & 0.66 & 0.61 & 0.55 & 0.56 & 0.49 & 0.43 & 0.60 & 0.60 & 0.57 & 0.60 \\ 
  Slovenia & 1.09 & 0.84 & 0.94 & 1.00 & 1.16 & 1.07 & 1.07 & 1.06 & 1.00 & 1.01 & 0.99 & 1.05 & 0.94 & 0.97 & 0.91 & 0.91 \\ 
  Spain & 0.57 & 0.52 & 0.61 & 0.54 & 0.58 & 0.62 & 0.74 & 0.68 & 0.70 & 0.68 & 0.63 & 0.59 & 0.53 & 0.54 & 0.56 & 0.57 \\ 
  Sweden & 0.83 & 0.87 & 0.89 & 0.80 & 0.85 & 0.83 & 0.88 & 0.84 & 0.86 & 0.82 & 0.70 & 0.67 & 0.62 & 0.71 & 0.77 & 0.61 \\ 
\bottomrule
\end{tabular}
\end{center}
\end{table}
\end{landscape}

\subsection{Independence of Irrelevant Alternatives}
\label{sec:iia}

As discussed in Sections \ref{sec:intro} and \ref{sec:pluce}, DEA rankings do not adhere to the independence of irrelevant alternatives (IIA) assumption of the Plackett–Luce distribution. This means, among other things, that if a DMU is removed from the set, the ranking of the remaining DMUs can change. The question is to what extent the IIA assumption is violated in real data.

We conduct a simple experiment. We remove a single DMU from the set, compute DEA efficiency, and compare the resulting ranking with the original ranking based on the full set of DMUs. We repeat this process for all DMUs and time periods. Thus, we obtain a total of $N \cdot T = 432$ DEA rankings. In total, 87 percent of the rankings remain unchanged after removing a single DMU. The correlation coefficient between the rankings based on the reduced sets and the full set is 0.9954. Naturally, the ranking is more likely to change when DMUs with higher efficiency scores are removed. In our case, removing countries such as Bulgaria, Cyprus, Luxembourg, Netherlands, Romania, and Slovenia---all of which hold high ranks according to Figure \ref{fig:interval}---results in changes to the ranking across multiple time periods. In summary, we confirm the violation of the IIA assumption in our empirical study. Nevertheless, the extent of this violation is rather mild, as evidenced by the relatively high correlation between rankings.

\subsection{Long-Term Ranking}
\label{sec:rank}

\begin{figure}
\begin{center}
\includegraphics[width=15cm]{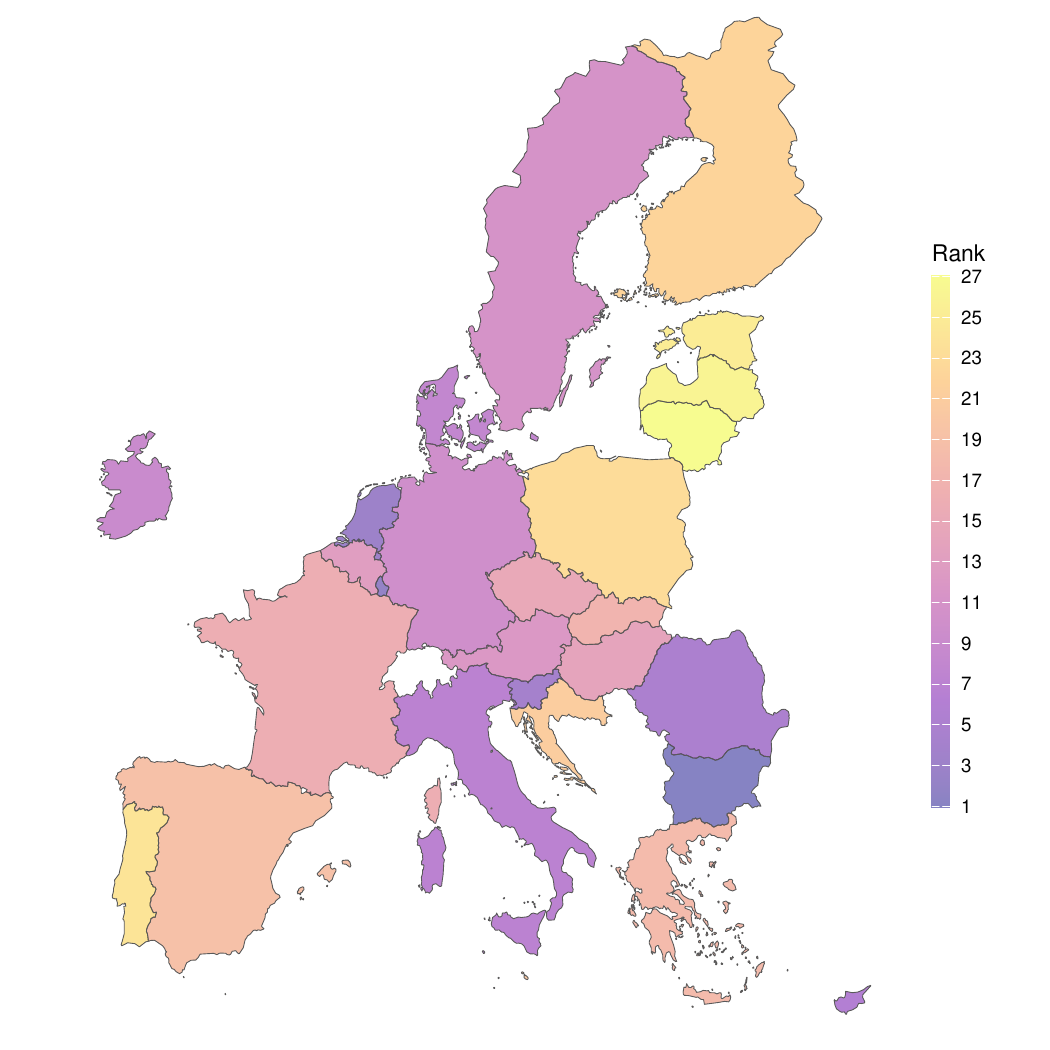}
\caption{The long-term ranking according to the dynamic ranking model.}
\label{fig:ranking}
\end{center}
\end{figure}

When conducting efficiency analysis using DEA across multiple time periods, it can be beneficial to report the long-term behavior. This could be done by simple aggregate statistics, such as modal or median ranks. But it is also a perfect task for our dynamic ranking model. For this purpose, we estimate the model without any contextual variables, only in the form of a stationary time series model. We can then rank DMUs according to the unconditional values of the worth parameters, which are simply equal to $\omega_{n}$. This long-term or ``ultimate'' ranking is visualized in Figure \ref{fig:ranking}.

\subsection{Panel Regression and Ranking Model}
\label{sec:panel}

We proceed to the second stage where we find relation between the efficiency scores or their associated rankings and the contextual variables. For the efficiency scores, we employ standard panel linear regression model with the robust estimation of the standard errors by the White method. As dependent variable, we use the efficiency scores obtained by the basic DEA model \eqref{eq:charnes} of \cite{Charnes1978} (denoted as CCR), the super-efficiency model \eqref{eq:andersen} of \cite{Andersen1993} (denoted as AP), and the universal DEA model \eqref{eq:hladik} of \cite{Hladik2019} (denoted as H). We also use the log transform of the AP efficiency scores, which are equal to the logit transform of H efficiency scores,
\begin{equation}
\theta^{Log}_{nt} = \ln \left( \theta^{AP}_{nt} \right) = - \ln \left( \frac{2}{\theta^H_{nt}} - 1 \right).
\end{equation}
Furthermore, we use the AP efficiency scores, or equivalently the H efficiency scores, to derive rankings of the DMUs, which serve as the dependent variable in our dynamic ranking model.

The results of the estimated models are reported in Table \ref{tab:panel}. All panel linear regression models exhibit consistent signs of coefficients. The only exception is the Rule of Law indicator, with a positive coefficient for the Log model but negative for the CCR, AP, and H models. All these coefficients are, however, very close to zero and statistically insignificant. In contrast, all panel linear regression models find the Voice and Accountability indicator to be statistically significant at the 0.05 level. Furthermore, the Government Effectiveness indicator is significant according to all panel regression models but AP. All the remaining contextual variables are found insignificant by all models.

The dynamic ranking model confirms the positive and significant relation to the Voice and Accountability indicator, which is consistent with the results of all panel regression models. However, regarding the Government Effectiveness indicator, the model agrees with AP and finds it to be insignificant. The Political Stability and GDP variables have opposite signs, in contrast to the panel regression models, but remain insignificant. It is important to note that while the signs and significance of coefficients can be compared between the panel regression models and the dynamic ranking model, the estimated values cannot be directly compared due to differences in the model specifications. The coefficients $\varphi$ and $\alpha$, which control the dynamics, have both positive values, as expected. The estimated value of 0.86 for $\varphi$ suggests that the process is stationary, with high persistence over time.

The inference of the ranking model is derived using the empirical Hessian (denoted as Hess. in Table \ref{tab:panel}). To ensure the robustness of our findings, we additionally employ the parametric bootstrap technique to compute standard errors and p-values (denoted as Boot. in Table \ref{tab:panel}). Our bootstrap procedure is based on 1\,000 simulated samples. According to Table \ref{tab:panel}, the estimated standard errors across the two methods are quite similar. An exception can be seen in the case of the Rule of Law variable; the bootstrapped standard deviation is noticeably lower here. Despite this discrepancy, the coefficient for this variable remains statistically insignificant at typical significance levels. The p-values exhibit similar behavior, and the significance of the variables remains unchanged; only the Voice and Accountability variable achieves significance at a lower level. Collectively, these findings affirm the validity of inference based on the empirical Hessian within our finite sample.

\begin{table}
\begin{center}
\caption{The estimated coefficients with standard errors for panel linear regressions and the dynamic ranking model.}
\label{tab:panel}
\begin{tabular}{lcccccc} 
\toprule
& \multicolumn{4}{c}{Efficiency Score} & \multicolumn{2}{c}{Rank} \\ \cmidrule(l{3pt}r{3pt}){2-5} \cmidrule(l{3pt}r{3pt}){6-7}
& CCR & AP & H & Log & Hess. & Boot. \\
\midrule
\multirow{2}{*}{Voice and Accountability} & $0.21^{*}$ & $0.45^{*}$ & $0.24^{*}$ & $0.52^{*}$ & $3.61^{**}$ & $3.61^{***}$ \\
& (0.09) & (0.20) & (0.10) & (0.21) & (1.21) & (1.21) \\
\\
\multirow{2}{*}{Political Stability} & $0.04$ & $0.10$ & $0.05$ & $0.11$ & $-0.60$ & $-0.60$ \\
& (0.07) & (0.09) & (0.06) & (0.16) & (0.56) & (0.61) \\
\\
\multirow{2}{*}{Government Effectiveness} & $0.23^{**}$ & $0.14$ & $0.17^{*}$ & $0.39^{*}$ & $0.65$ & $0.65$ \\
& (0.08) & (0.14) & (0.08) & (0.19) & (0.63) & (0.68) \\
\\ 
\multirow{2}{*}{Regulatory Quality} & $-0.09$ & $-0.15$ & $-0.08$ & $-0.19$ & $-1.21$ & $-1.21$ \\
& (0.10) & (0.15) & (0.09) & (0.20) & (0.69) & (0.74) \\
\\ 
\multirow{2}{*}{Rule of Law} & $-0.03$ & $-0.02$ & $-0.01$ & $0.02$ & $0.02$ & $0.02$ \\
& (0.14) & (0.19) & (0.12) & (0.27) & (0.92) & (0.25) \\
\\
\multirow{2}{*}{Control of Corruption} & $-0.08$ & $-0.14$ & $-0.08$ & $-0.20$ & $-0.75$ & $-0.75$ \\
& (0.11) & (0.19) & (0.11) & (0.24) & (0.75) & (0.74) \\
\\
\multirow{2}{*}{Gross Domestic Product} & $-0.02$ & $-0.23$ & $-0.09$ & $-0.19$ & $1.08$ & $1.08$ \\
& (0.25) & (0.28) & (0.19) & (0.42) & (1.59) & (1.64) \\
\\
\multirow{2}{*}{Autoregressive Parameter $\varphi$} & & & & & $0.86^{***}$ & $0.86^{***}$ \\
& & & & & (0.06) & (0.10) \\
\\ 
\multirow{2}{*}{Score Parameter $\alpha$} & & & & & $0.96^{***}$ & $0.96^{***}$ \\
& & & & & (0.08) & (0.10) \\
\\ 
\bottomrule
\multicolumn{7}{r}{\textit{Note:} $^{***}p < 0.001$; $^{**}p < 0.01$; $^{*}p < 0.05$}
\end{tabular}
\end{center}
\end{table}

\subsection{Computing Environment}
\label{sec:comp}

The empirical study was performed in R. The CCR and AP DEA efficiency scores were obtained using the \verb"dea()" and \verb"sdea()" functions from the \verb"Benchmarking" package. The H efficiency scores were obtained from the AP DEA efficiency scores using transformation \eqref{eq:transform}. The panel regressions were estimated using the \verb"plm()" function from the \verb"plm" package with robust inference obtained using the \verb"coeftest()" function from the \verb"lmtest" package. The dynamic ranking model was estimated by the \verb"gas()" and \verb"gas_bootstrap()" functions from the \verb"gasmodel" package. All these packages are available on CRAN.

\subsection{Discussion of Results}
\label{sec:res}

The results of our analysis show that the Voice and Accountability indicator has a consistently positive and significant correlation with research efficiency across all models. This indicates that factors such as participation in selecting the government, freedom of expression, freedom of association, and freedom of media, which form the Voice and Accountability indicator, play a crucial role in enhancing research efficiency. In contrast, the Government Effectiveness indicator also has a positive effect on research efficiency, but its significance is not confirmed by all models. This suggests that while Government Effectiveness can enhance research efficiency, it may not be as crucial as the Voice and Accountability indicator and lacks robustness. The findings of this study can inform policy decisions and strategic planning to enhance research performance and impact, ultimately advancing knowledge and innovation in various fields.

\section{Conclusion}
\label{sec:con}

The main contribution of the paper lies in proposing and showcasing the use of the dynamic ranking model of \cite{Holy2022f} in the second stage of DEA. The primary objective of the model is to serve as a complement to conventional second-stage models and provide a robustness check. In the empirical study evaluating research efficiency in the higher education sector, we find that the results of the second stage may slightly differ when using a panel regression model applied to the efficiency scores from various DEA models and when using the studied dynamic ranking model. All models agree that the Voice and Accountability variable, one of the indicators of good governance, has a positive and significant impact on research efficiency. Additionally, all models suggest that the Political Stability, Regulatory Quality, Rule of Law, and Control of Corruption indicators do not significantly correlate with research efficiency. However, the results on the Government Effectiveness indicator are mixed---while some models suggest that it has a significant positive effect, others, including the dynamic ranking model, consider this variable insignificant. As the second stage of DEA is surrounded by controversies about the appropriateness of its use, while being widely applied by researchers at the same time, it is important to perform various robustness checks and not rely on results from a single model. In this sense, the dynamic ranking model can be quite useful as it is fundamentally different from traditional panel models and is thus able to provide a different perspective. However, just like traditional second-stage models, it also suffers from the drawback of misspecifying the process generating the efficiency scores and rankings. It can also be used only for repeated rankings according to efficiency scores, such as annual assessment of efficiency, that do not include ties. While the use of the dynamic ranking model in the second stage of DEA may not be a perfect solution for all situations, it is a valuable addition to the DEA researcher's toolkit.

As an additional contribution, we show that the recently proposed universal DEA model of \cite{Hladik2019}, with attractive properties and interpretation, is closely related to the widely used super-efficiency DEA model of \cite{Andersen1993}. Specifically, we demonstrate that the efficiency scores from one model can be derived from the other model using a simple transformation.

Future research efforts should be directed towards expanding the dynamic ranking model in two ways. Firstly, the model should be able to incorporate ties, which may occur due to DEA models lacking super-efficiency. Secondly, the model should be able to capture more complex interdependencies between DMUs, which can be perhaps achieved or at least approximated by employing Thurstone order statistics models based on the multivariate normal or multivariate extreme value distributions.

\section*{Acknowledgements}
\label{sec:acknow}

I would like Jan Zouhar for his comments. Computational resources were supplied by the project "e-Infrastruktura CZ" (e-INFRA LM2018140) provided within the program Projects of Large Research, Development and Innovations Infrastructures.

\section*{Funding}
\label{sec:fund}

The work on this paper was supported by the Czech Science Foundation under project 23-06139S and the personal and professional development support program of the Faculty of Informatics and Statistics, Prague University of Economics and Business.


\end{document}